\documentclass[preprint,showpacs,preprintnumbers,amsmath,amssymb]{revtex4}
\usepackage{graphicx}% Include figure files
\usepackage{dcolumn}% Align table columns on decimal point
\usepackage{bm}% bold math
\newcommand{\comment}[1]{}
\newcommand{\bea}{\begin{eqnarray}}
\newcommand{\eea}{\end{eqnarray}}

\begin{document}
%\eqnobysec

\title[Factorization technique for Coupled Li\'enard Equations]
{Exact solutions of coupled Li\'enard-type nonlinear systems using factorization technique}

\author{Tamaghna Hazra$^1$, V. K. Chandrasekar$^2$, R Gladwin Pradeep$^2$ and M. Lakshmanan$^2$}
\affiliation{$^1$Department of Physics, Indian Institute of Technology Kanpur, Kanpur 208016,India \\ $^2$Centre for Nonlinear Dynamics, Department of Physics, 
Bharathidasan University, Tiruchirappalli - 620 024, India}

\date{\today} 

\begin{abstract}
General solutions of nonlinear ordinary differential equations (ODEs) are in general difficult to find although powerful integrability techniques exist in the literature for this purpose. It has been shown that in some scalar cases particular solutions may be found with little effort if it is possible to factorize the equation in terms of first order differential operators. In our present study we use this factorization technique to address the problem of finding solutions of a system of general two-coupled Li\'enard type nonlinear differential equations. We describe a generic algorithm to identify specific classes of Li\'enard type systems for which solutions may be found. We demonstrate this method by identifying a class of two-coupled equations for which the particular solution can be found  by solving a Bernoulli equation. This class of equations include coupled generalization of the modified Emden equation. We further deduce the general solution of a class of coupled ordinary differential equations using the factorization procedure discussed in this manuscript. 
 
\end{abstract}

\maketitle
\section{Introduction}
\label{sec1}
The modified Emden equation (MEE)
\begin{align}
\ddot{u}+\alpha u\dot{u}+\beta u^3=0,\label{mee}
\end{align}
where overdot denotes differentiation with respect to time, has been studied by Painlev\'e \cite{painleve} more than a century ago. He and his coworkers found that \eqref{mee} is integrable for two specific parametric choices, namely (i) $\beta=-\alpha^2$ and (ii) $\beta=\frac{\alpha^2}{9}$. Recently Chandrasekar \emph{et al} \cite{gensoln} have shown that this equation is integrable for any choice of arbitrary parameters $\alpha$ and $\beta$. The same authors have also underlined its immense applicability through its intimate connections with other well-known systems like the force-free Duffing-type oscillator and the two-dimensional Lotka-Volterra system. It also has substantial importance in physics itself as it rears its head in a plethora of varied contexts such as the equilibrium configurations of a spherical gas cloud acting under the mutual attraction of its molecules and subject to the laws of thermodynamics \cite{gas11,gas12,gas21,gas22}, and in the modeling of fusion of pellets \cite{fusion}. It is also the governing equation for the spherically symmetric expansion or collapse of a relativistically gravitating mass \cite{stellar1,stellar2,stellar3}. Equation \eqref{mee} can also be seen as a one-dimensional analog of the boson `gauge-theory' equations introduced by Yang and Mills \cite{boson1,boson2}. This equation also comes up in a variety of mathematical problems such as univalued functions defined by second-order differential equations \cite{math1} and the Riccati equation \cite{boson1}.

Equation \eqref{mee} is a particular case of a more general family of equations of the Li\'enard type,
\begin{align}
\ddot{u}+f(u)\dot{u}+g(u)=0.\label{lienard}
\end{align}
The  Li\'enard family of equations is in general not integrable, and only for some specific choices of the arbitrary functions $f(u)$ and $g(u)$ complete solutions can be found. Its applications and properties have been discussed in \cite{idunno1,idunno2}. However, in general, the task of finding general solutions of such nonlinear second order systems requires the use of complicated, albeit powerful, procedures like the Painlev\'e analysis, modified Prelle-Singer procedure, use of Lax pairs and the associated inverse scattering transforms. Recently Rosu and Cornejo-P\'erez \cite{cornejo2,cornejo2} have suggested that at least in the case of some polynomial nonlinearities particular solutions may be found rather simply by an elegant method of factorizing the system \eqref{lienard} into two first order differential operators. Berkovich \cite{berk} has explored the factorization technique for solving scalar ODEs in substantial detail. The factorizing procedure is also used to obtain the travelling wave solutions of certain nonlinear partial differential equations  such as the Korteweg-de Vries-Burgers equation, Camassa-Holm equation, Kolmogorov-Petrovskii-Piskunov equation \cite{cornejo2,li}. However this approach has not yet been applied to the study of coupled systems. Our motivation in this present study is to extend this approach to the study of coupled systems and identify which systems allow solutions to be found by such means. 

In our present paper we study a system of two-coupled Li\'enard type equations,
\begin{subequations}
\label{cpl}
\begin{eqnarray}
&&\ddot{u}+f_1(u,v)\dot{u}+f_2(u,v)\dot{v}+g_1(u,v)=0,\label{cpl1}\\
&&\ddot{v}+f_3(u,v)\dot{u}+f_4(u,v)\dot{v}+g_2(u,v)=0,\label{cpl2}
\end{eqnarray}
\end{subequations}
and investigate whether specific forms of this generic equation allow solutions to be found by the factorization technique. Here we show that this can be done by factorizing the system (\ref{cpl}) in terms of first-order differential operators by generalizing the procedure of Reyes and Rosu \cite{scfact} for scalar second-order ODEs. We find that for the following general class of equations finding a solution is as simple as solving a Bernoulli equation in one variable:
\begin{subequations}
\label{ultragen}
\begin{align}
\ddot{u}-b_2\eta \dot{\psi_1}+b_1\xi \dot{\psi_2}+(b_1\psi_2-b_2\psi_1) \dot{\bar{h}} &+(b_1\dot{\psi}_2-b_2\dot{\psi}_1) \bar{h} +\delta[b_2(\eta-d_1)(\bar{h}+d_1)\psi_1\notag\\
											     &-b_1(\xi -d_2)(\bar{h}+d_2)\psi_2]=0,\\
\ddot{v}+a_2\eta \dot{\psi_1}-a_1\xi \dot{\psi_2}-(a_1\psi_2-a_2\psi_1) \dot{\bar{h}}-&(a_1\dot{\psi}_2-a_2\dot{\psi}_1) \bar{h} -\delta[a_2(\eta-d_1)(\bar{h}+d_1)\psi_1\notag\\
											     &-a_1(\xi -d_2)(\bar{h}+d_2)\psi_2]=0,
\end{align}
\end{subequations}
where $\psi_{1,2}=a_{1,2}u+b_{1,2}v+c_{1,2}$, $\eta, \xi$ are arbitrary functions of $u$ and $v$, $\bar{h}$ is an arbitrary \textit{homogeneous} function in $\psi_1$ and $\psi_2$, $\delta=a_1b_2-a_2b_1$ and $d_{1,2},a_{1,2},b_{1,2},c_{1,2}$ are all constants. For suitable choice of arbitrary functions and parameters, namely $\eta=\bar{h}, \quad \xi=\bar{h}, \quad \psi_1=u, \quad \psi_2=v \mbox{ and } \bar{h}=-(k_1u+k_2v)$, where $k_1$ and $k_2$ are constants, this reduces to the form
\begin{subequations}
\label{cMEE}
\begin{align}
\ddot{u}+2(k_1u+k_2v)\dot{u}+(k_1\dot{u}+k_2\dot{v})u+(k_1u+k_2v)^2u+\omega_1^2u=0,  \\
\ddot{v}+2(k_1u+k_2v)\dot{v}+(k_1\dot{u}+k_2\dot{v})v+(k_1u+k_2v)^2v+\omega_2^2v=0,
\end{align}
\end{subequations}
where $\omega_{1,2}=d_{1,2}$. This system has been studied in the literature \cite{nMEE} and shown to be completely integrable by Chandrasekar \textit{et al} \cite{cMEE}. Also \eqref{cMEE} is a two-coupled version of the special modified Emden type equation with additional linear forcing,
\begin{align}
\ddot{u}+3k u\dot{u}+k^2 u^3+\omega^2 u=0.\label{meeforc}
\end{align}
This equation exhibits the isochronous property \cite{isochr}. 

The plan of the paper is as follows. In Section \ref{sec2} we investigate the factorization technique for the two-coupled system (\ref{cpl}) in terms of first order differential operators. In section \ref{sec1.5} we introduce a generic algorithm to identify classes of coupled Li\'enard type equations (\ref{cpl}) whose solutions can be found by simple methods and chalk out the procedure to identify a class whose solution may be found in terms of the Bernoulli equation. In section \ref{sec4.5} we discuss the procedure to deduce the general solution of a class of coupled ODEs using the factorization procedure.  We illustrate the procedure by considering a specific equation. Finally in section \ref{sec6} we conclude by summarizing our main results.  In  appendix A we prove that the functions $\psi_{1,2}$ in Eqs. (\ref{ultragen}) should be linear in $u$ and $v$.  In appendix B, we discuss the factorization procedure for the equations (\ref{redgen}) in which $h_1$ is a function of $\psi_1$ alone and $h_2$ is a function of $\psi_2$ only.  In appendix C, we show that the procedure to obtain the general solution for coupled ODEs discussed in this paper can also be straightforwardly used to obtain the general solutions of the scalar equations discussed in Ref. \cite{cornejo2}.

\section{Factorization of the general case}
\label{sec2}

If the coupled Li\'enard system (\ref{cpl1}-\ref{cpl2}) can be factorized in the form:
\begin{subequations}
\label{fact}
\begin{eqnarray}
&&[D-\phi_1(u,v)][D-\phi_2(u,v)]\psi_1(u,v)=0,\label{fact1}\\ 
&&[D-\phi_3(u,v)][D-\phi_4(u,v)]\psi_2(u,v)=0,\label{fact2}
\end{eqnarray}
\end{subequations}
where $D=\frac{d}{dt}$, $\psi_{1,2}$ are arbitrary functions of $u$ and $v$, and $\phi_i$ are functions of $u$ and $v$ to be determined, then the problem of finding the general solution of (\ref{cpl}) may be addressed by simultaneously solving the following set of first order differential equations:
\begin{subequations}
\label{solgen}
\begin{align}
[D-\phi_1(u,v)]P_1(u,v)&=0,\\
[D-\phi_3(u,v)]P_2(u,v)&=0,\\
[D-\phi_2(u,v)]\psi_1(u,v)&=P_1(u,v),\\
[D-\phi_4(u,v)]\psi_2(u,v)&=P_2(u,v).
\end{align}
\end{subequations}
Also, as pointed out by Rosu and Cornejo-P\'erez \cite{cornejo2} for the scalar case, the problem of finding particular solutions to Eq. \eqref{cpl} can be addressed by solving the reduced equations, 
\begin{subequations}
\label{red}
\begin{eqnarray}
&&[D-\phi_2(u,v)]\psi_1(u,v)=0,\label{red1}\\
&&[D-\phi_4(u,v)]\psi_2(u,v)=0\label{red2},
\end{eqnarray}
\end{subequations}
which may in some cases be relatively simple.

We are thus motivated to explore the question: \emph{For what forms of $f_i$ and $g_i$ in (\ref{cpl}) can the system of coupled equations be explicitly factorized in the form (\ref{fact})?}. To answer this question we proceed as follows.

By expanding (\ref{fact}) and manipulating the resulting equations appropriately, we get the equivalent of (\ref{cpl}) as
\begin{subequations}
\label{cplfact1}
\begin{align}
\ddot{u}+&\frac{1}{\delta}[(\psi_{1uu}\psi_{2v}-\psi_{2uu}\psi_{1v})\dot{u}^2+(\psi_{1uv}\psi_{2v}-\psi_{2uv}\psi_{1v})\dot{u}\dot{v}+(\psi_{1vv}\psi_{2v}-\psi_{2vv}\psi_{1v})\dot{v}^2]\notag \\
&+\tilde{f_1}\dot{u}+\tilde{f_2}\dot{v}+\tilde{g_1}=0,&  \\
\ddot{v}+&\frac{1}{\delta}[(\psi_{1uu}\psi_{2u}-\psi_{2uu}\psi_{1u})\dot{u}^2+(\psi_{1uv}\psi_{2u}-\psi_{2uv}\psi_{1u})\dot{u}\dot{v}+(\psi_{1vv}\psi_{2u}-\psi_{2vv}\psi_{1u})\dot{v}^2]\notag \\
&+\tilde{f_3}\dot{u}+\tilde{f_4}\dot{v}+\tilde{g_2}=0,&\label{cplfact}
\end{align}
\end{subequations}
where the functions
\begin{subequations}
\label{fdef}
\begin{align}
\tilde{f_1}&=\frac{1}{\delta}[(\phi_3+\phi_4)\psi_{2u}\psi_{1v}- (\phi_1+\phi_2)\psi_{1u}\psi_{2v} -\phi_{2u}\psi_1\psi_{2v} + \phi_{4u}\psi_2\psi_{1v}], \\
\tilde{f_2}&= \frac{1}{\delta}[ (\phi_3+\phi_4-\phi_1-\phi_2)\psi_{1v}\psi_{2v} -\phi_{2v}\psi_1\psi_{2v} + \phi_{4v}\psi_2\psi_{1v}], \\
\tilde{f_3}&=\frac{1}{\delta}[\phi_{2u}\psi_1\psi_{2u} -(\phi_3+\phi_4-\phi_1-\phi_2)\psi_{1u}\psi_{2u} - \phi_{4u}\psi_2\psi_{1u}],  \\
\tilde{f_4}&=\frac{1}{\delta} [ (\phi_1+\phi_2)\psi_{2u}\psi_{1v} - (\phi_3+\phi_4)\psi_{1u}\psi_{2v} +\phi_{2v}\psi_1\psi_{2u} - \phi_{4v}\psi_2\psi_{1u}], 
\end{align}
\end{subequations}
\begin{subequations}
\begin{align}
\tilde{g_1}&=\frac{1}{\delta}\left[\phi_1\phi_2\psi_{2v}\psi_1 - \phi_3\phi_4\psi_{1v}\psi_2\right],  \\
 \tilde{g_2}&=\frac{1}{\delta}\left[\phi_3\phi_4\psi_{1u}\psi_2-\phi_1\phi_2\psi_{2u}\psi_1 \right], \label{gdef}
\end{align}
and the quantity
\begin{align}
\delta&=\psi_{1u}\psi_{2v}-\psi_{2u}\psi_{1v}\neq 0.& \label{delta}
\end{align}
\end{subequations}

Comparing Eq. (\ref{cpl}) with Eq. (\ref{cplfact1}),  we find $\psi_{1,2uu}=\psi_{1,2uv}=\psi_{1,2vv}=0$.  This implies that one can assume 
\begin{align}
\psi_{1,2}=a_{1,2}u+b_{1,2}v+c_{1,2},\label{psidef}
\end{align}
without loss of generality. Here $a_i,\,b_i,$ and $,c_i$ are arbitrary constants.  Using \eqref{psidef} in \eqref{delta}, we infer that $\delta=a_1b_2-a_2b_1=\mbox{constant}$.  Further, defining the variables $f_i=\delta\tilde{f}_i,\,i=1,2,3,4$ and $g_j=\delta\tilde{g}_j$, $j=1,2,$ we obtain the relations
\begin{align}
&a_1g_1+b_1g_2=\phi_1\phi_2\psi_1, &a_2g_1+b_2g_2=\phi_3\phi_4\psi_2.\label{ggen2}
\end{align}

From the above two equations in (\ref{ggen2}) one can solve for $\phi_1$ and $\phi_3$ as
\begin{align}
&\phi_1=\frac{1}{\phi_2}\bigg[\frac{a_1g_1+b_1g_2}{\psi_1} \bigg], \;\;
&\phi_3=\frac{1}{\phi_4}\bigg[\frac{a_2g_1+b_2g_2}{\psi_2} \bigg]. \label{phi13gen2}
\end{align}
Using \eqref{phi13gen2} to eliminate $\phi_1$ and $\phi_3$ in (\ref{fdef}) one can, after some manipulations, get a set of PDEs which determine $\phi_2$ and $\phi_4$:
\begin{align}
\frac{\partial}{\partial u}\left[\frac{\phi_2^2\psi_1^2}{2}\right]=-a_1(a_1g_1+b_1g_2)-(a_1f_1+b_1f_3)\phi_2\psi_1,\label{pde1}\\
\frac{\partial}{\partial v}\left[\frac{\phi_2^2\psi_1^2}{2}\right]=-b_1(a_1g_1+b_1g_2)-(a_1f_2+b_1f_4)\phi_2\psi_1,\label{pde2}\\
\frac{\partial}{\partial u}\left[\frac{\phi_4^2\psi_2^2}{2}\right]=-a_2(a_2g_1+b_2g_2)-(a_2f_1+b_2f_3)\phi_2\psi_1,\label{pde3}\\
\frac{\partial}{\partial v}\left[\frac{\phi_4^2\psi_2^2}{2}\right]=-b_2(a_2g_1+b_2g_2)-(a_2f_2+b_2f_4)\phi_2\psi_1.\label{pde4}
\end{align}
Compatibility of \eqref{pde1} with \eqref{pde2} and of \eqref{pde3} with \eqref{pde4} requires that
\begin{align}
\frac{\partial}{\partial v}\left[a_1(a_1g_1+b_1g_2)+(a_1f_1+b_1f_3)\phi_2\psi_1\right]&=\frac{\partial}{\partial u}\left[b_1(a_1g_1+b_1g_2)+(a_1f_2+b_1f_4)\phi_2\psi_1\right],\notag\\
\frac{\partial}{\partial v}\left[a_2(a_2g_1+b_2g_2)+(a_2f_1+b_2f_3)\phi_2\psi_1\right]&=\frac{\partial}{\partial u}\left[b_2(a_2g_1+b_2g_2)+(a_2f_2+b_2f_4)\phi_2\psi_1\right],\notag
\end{align}
which implies
\begin{subequations}
\label{compatab}
\begin{align}
&\psi_1 \hat{f_2}\frac{\partial\phi_{2}}{\partial u}-\psi_1 \hat{f_1}\frac{\partial \phi_{2}}{\partial v}+ (a_1\hat{f_2}-b_1\hat{f_1} + [\hat{f_{2u}}-\hat{f_{1v}}]\psi_1)\phi_2 + b_1\hat{g_{1u}} - a_1\hat{g_{1v}}=0,\label{compata}\\
&\psi_2\hat{f_4}\frac{\partial \phi_{4}}{\partial u}-\psi_2\hat{f_3}\frac{\partial \phi_{4}}{\partial v} + (a_2\hat{f_4}-b_2\hat{f_3} + [\hat{f_{4u}}-\hat{f_{3v}}]\psi_2)\phi_4 + b_2\hat{g_{2u}} - a_2\hat{g_{2v}}=0,\label{compatb}
\end{align}
\label{compat}
\end{subequations}
where
\begin{subequations}
\label{hatgf}
\begin{align}
&\hat{g_1}=a_1g_1+b_1g_2, &\hat{g_2}=a_2g_1+b_2g_2,\label{hatg}\\
&\hat{f_1}=a_1f_1+b_1f_3, &\hat{f_2}=a_1f_2+b_1f_4,\\
&\hat{f_3}=a_2f_1+b_2f_3, &\hat{f_4}=a_2f_2+b_2f_4.\label{hatf}
\end{align}
\end{subequations}
Note that in Eq. (\ref{hatgf}), the various functions $f_i$s, $i=1,2,3,4$ and $g_i$s, $i=1,2$, are just the coefficients occurring in the coupled ODE (\ref{cpl}).

We thus have two first order linear PDEs (\ref{compata}) and (\ref{compatb}) for $\phi_2$ and $\phi_4$, respectively, which may be solved to find the explicit forms of $\phi_2$ and $\phi_4$. When this is found one can find the explicit forms of $\phi_1$ and $\phi_3$ from \eqref{phi13gen2}, which imply that the factorization (\ref{fact})  is complete.

Thus we have answered the question posed at the beginning of this section. The system of coupled Li\'enard type equations (\ref{cpl}) can be factorized in the form (\ref{fact}) if and only if the pair of first order PDEs\eqref{compat} can be solved explicitly for $\phi_2$ and $\phi_4$ in terms of $u$ and $v$. Conversely, for every solution $\phi_2$ and $\phi_4$ for (\ref{compat}), there exists a factorization (\ref{fact})) corresponding to a system of coupled second order nonlinear ODEs of the form (\ref{cpl}).

\section{Outline of the procedure}
\label{sec1.5}

We next turn to the problem of finding solutions by using the factorized form(\ref{fact}). If the lower order forms \eqref{solgen} and (\ref{red}) can be associated with some known system of equations for which the solution is known, then the solution of the original coupled second order system (\ref{cpl}) can be found. With this motivation we explore the inverse problem: \emph{Given some particular forms of $\phi_i$, $i=1,2,3,4,$ satisfying Eqs. (\ref{phi13gen2}) and (\ref{compatab}), what is the subclass of two-coupled Li\'enard type systems (\ref{cplfact1}) that can be factorized to the form (\ref{solgen}) and (\ref{red}), for some choice of arbitrary constants $a_i,b_i,c_i$?}

We start off by identifying such an interesting choice of $\phi_i$.\\ 
Rewriting (\ref{red}) we have
\begin{align}
&\dot{\psi_1}=\phi_2\psi_1, \:\:\dot{\psi_2}=\phi_4\psi_2 \label{expred}
\end{align}

We consider the case that the reduced equations are of the form
\begin{subequations}
\label{redgen}
\begin{align}
\dot{\psi_1}=(\omega_1+h_1(\psi_1,\psi_2))\psi_1,\label{redgen1} \\
\dot{\psi_2}=(\omega_2+h_2(\psi_1,\psi_2))\psi_2, \label{redgen2} 
\end{align}
\end{subequations}
where $\omega_1$ and $\omega_2$ are constants, so that $\phi_2=\omega_1+h_1(\psi_1,\psi_2),\:\:\phi_4=\omega_2+h_2(\psi_1,\psi_2)$.
If $h_1$ is independent of $\psi_2$ and $h_2$ is independent of $\psi_1$, then the reduced equations are uncoupled and the problem of finding particular solutions reduces to the much simpler problem of solving a pair of scalar first order ODEs. An example of this is dealt with in Appendix \ref{app2}.
 
Otherwise if $h_1=\alpha h_2$ where $\alpha$ is a constant then one integral can be found by eliminating $h_1$ from (\ref{redgen1}) and (\ref{redgen2}) as
\begin{align}
\alpha \dot{\psi_1}\psi_2-\dot{\psi_2}\psi_1&=(\alpha \omega_1-\omega_2)\psi_1 \psi_2.\label{rednxt}
\end{align}
Eq.\eqref{rednxt} can be rearranged to
\begin{align}
 \qquad \qquad \qquad \alpha \frac{\dot{\psi_1}}{\psi_1}-\frac{\dot{\psi_2}}{\psi_2}&=(\alpha \omega_1-\omega_2).\label{rednx2}
\end{align}
Integrating Eq.\eqref{rednx2} one gets
\begin{align}
\psi_1^\alpha=C_1e^{(\alpha \omega_1-\omega_2)t}\psi_2 \equiv c(t) \psi_2,\label{cdef}
\end{align}
where $C_1$ is the integration constant, $c(t)=C_1e^{(\alpha \omega_1-\omega_2)t}$. 

Let us now consider the specific case $h_1=\alpha h_2=h(\psi_1,\psi_2)$, where $h(\psi_1,\psi_2)$ is a polynomial containing $N$ terms and is of the form
\begin{eqnarray}
h(\psi_1,\psi_2)=\sum_{i=1}^Nk_i\psi_1^{p_i}\psi_2^{q_i},\label{hdef3}
\end{eqnarray}
where $p_i$ and $q_i$ are real constants.  Using the relation (\ref{cdef}) and rewriting the above equation we get
\begin{eqnarray}
h=\sum_{i=1}^N k_i  \left(c(t)\psi_2\right)^{\frac{p_i}{\alpha}}\psi_2^{q_i}.\label{hdef}
\end{eqnarray}
The above polynomial becomes homogeneous for the condition $\frac{p_i}{\alpha}+q_i=m$, where $m$ is a constant,
\begin{eqnarray}
h=\psi_2^{m}\sum_{i=1}^N k_i  c(t)^{\frac{p_i}{\alpha}}.\label{hdef2}
\end{eqnarray}

%Considering the case $\alpha=1$ \mbox{and} $h_1=h_2 \equiv h$ is homogeneous in $\psi_1$ and $\psi_2$ of order $q$ (say), we can write:
%\begin{align}
%&h(\psi_1,\psi_2)=k_0\psi_1^q+k_1\psi_1^{q-1}\psi_2+k_2\psi_1^{q-2}\psi_2^2+...+k_{q-1}\psi_1\psi_2^{q-1}+k_q\psi_2^q \label{hdef},\\
%&\qquad\qquad= k_0 [c(t)]^q \psi_2^q + k_1 [c(t)]^{q-1}\psi_2^q+...+k_q \psi_2^q\equiv \psi_2^q d(t),\label{hdef2}
%\end{align}
%where $d(t)=k_0 [c(t)]^q +k_1 [c(t)]^{q-1} + ... + k_{q-1} c(t) +k_q$.  
Next, substituting \eqref{psidef} in (\ref{cplfact1}) and simplifying we get
\begin{subequations}
\label{example1}
\bea
&&\hspace{-2cm}\ddot{u}+\frac{1}{\delta}\bigg[(a_2b_1\phi_3+\omega_2+h)-a_1b_2(\phi_1+\omega_1+h)+(b_1\psi_2-b_2\psi_1)h_u\bigg]\dot{u} \notag\\
 &&       +\frac{1}{\delta}\bigg[b_1b_2(\phi_3-\phi_1+\omega_2-\omega_1)            +(b_1\psi_2-b_2\psi_1)h_v\bigg]\dot{v} \notag\\
&& \hspace{1cm}  +\frac{1}{\delta}\bigg[b_2\phi_1(\omega_1+h)\psi_1-b_1\phi_3(\omega_2+h)\psi_2\bigg]=0, \\
&&\hspace{-2cm}\ddot{v}-\frac{1}{\delta}\bigg[a_1b_2(\phi_3+\omega_2+h)-a_2b_1(\phi_1+\omega_1+h)+(a_1\psi_2-a_2\psi_1)h_v\bigg]\dot{v} \notag\\
&&        -\frac{1}{\delta}\bigg[a_1a_2(\phi_3-\phi_1+\omega_2-\omega_1)
+(a_1\psi_2-a_2\psi_1)h_u\bigg]\dot{u} \notag\\
  &&\hspace{1cm}      -\frac{1}{\delta}\bigg[a_2\bar{g_1}(\omega_1+h)\psi_1-a_1\bar{g_2}(\omega_2+h)\psi_2\bigg]=0, \label{cplgen1}
\eea
\end{subequations}
where $h$ is given by \eqref{hdef2} and $\phi_1$ and $\phi_3$ are given by
\begin{align}
&\phi_1=\frac{a_1g_1+b_1g_2}{\psi_1\phi_2}, &\phi_3=\frac{a_2g_1+b_2g_2}{\psi_2\phi_4}. \label{Gdef}
\end{align}
Redefining now the constants and functions as $\omega_1=\delta d_1,\qquad \omega_2=\delta d_2,\qquad \phi_1=\delta(\eta-d_1),\qquad \phi_3=\delta(\xi-d_2),\qquad h=\delta\bar{h}$, where $\delta=a_1b_2-a_2b_1=$constant, we find that \eqref{example1} reduces to the coupled Eq. \eqref{ultragen}.
A particular solution of (\ref{example1}) can be obtained by solving the following Bernoulli type equation
\begin{align}
\dot{\psi_2}=\omega_2\psi_2 +d(t)\psi_2^{m+1},\label{bern}
\end{align}
where $d(t)=\sum_{i=1}^Nk_ic(t)^{\frac{p_i}{\alpha}}$,
 which is obtained by substituting (\ref{hdef2}) in Eq. (\ref{redgen2}).  Equation (\ref{bern}) admits the following explicit solution 
\begin{align}
\psi_2(u,v)=\frac{e^{\omega_2t}}
{\left[C_2-m \int e^{m\omega_2t}d(t) dt \right]^{-\frac{1}{m}}},\label{sol1}
\end{align}
where $C_2$ is an integration constant. and, since $\psi_1=\psi_2 c(t)$, we have
\begin{align}
\psi_1(u,v)=\frac{c(t)e^{\omega_2t}}
{\left[C_2-m \int e^{m\omega_2t}d(t) dt \right]^{-\frac{1}{m}}}\label{sol2}.
\end{align}
Inverting the relation (\ref{psidef}) one can obtain $u$ and $v$ as
\begin{subequations}
\label{cplsolnn}
\begin{align}
u=\frac{b_2}{\delta}(\psi_1-c_1)-\frac{b_1}{\delta}(\psi_2-c_2),\\
v=\frac{a_1}{\delta}(\psi_2-c_2)-\frac{a_2}{\delta}(\psi_1-c_1),\label{inv}
\end{align}
\end{subequations}
Using the explicit forms of $\psi_1$ and $\psi_2$ as given by  Eqs. (\ref{sol1}) and \eqref{sol2}, we can then write
\begin{eqnarray}
&&\hspace{-0.7cm}u=\frac{b_2}{\delta}\left(\frac{c(t)e^{\omega_2t}}
{\left[C_2-m \int e^{m\omega_2t}d(t) dt \right]^{-\frac{1}{m}}}-c_1\right)-\frac{b_1}{\delta}\left(\frac{e^{\omega_2t}}
{\left[C_2-m \int e^{m\omega_2t}d(t) dt \right]^{-\frac{1}{m}}}-c_2\right),\\
&&\hspace{-0.7cm}v=\frac{a_1}{\delta}\left(\frac{e^{\omega_2t}}
{\left[C_2-m\int e^{m\omega_2t}d(t) dt \right]^{-\frac{1}{m}}}-c_2\right)
-\frac{a_2}{\delta}\left(\frac{c(t)e^{\omega_2t}}
{\left[C_2-m \int e^{m\omega_2t}d(t) dt \right]^{-\frac{1}{m}}}-c_1\right),
\end{eqnarray}
where $C_1$ in $c(t)$ (vide Eq. (\ref{cdef})) and $C_2$ are arbitrary integration constants.

\subsection{An example : Case $\psi_1=u$, $\psi_2=v$}
\label{sec3.5}
We now consider a specific equation belonging to the class of coupled ODEs \eqref{example1} and obtain a particular solution using the above factorization procedure.
%\label{sec3.5a}
For illustrative purpose, let us consider the simple case $\psi_1=u$ and $\psi_2=v$ for which Eq. \eqref{cplgen1} reduces to the form
\begin{subequations}
\label{uvgen}
\begin{align}
\ddot{u}-u\dot{h}+[h+\omega_1+\phi_1]\dot{u}-h_vu\dot{v}+&\phi_1u(h+\omega_1)=0,  \\
\ddot{v}-h_uv\dot{u}+v\dot{h}+[h+\omega_2+\phi_3]\dot{v}+&\phi_3v(h+\omega_2)=0. 
\end{align}
\end{subequations}
Choosing the arbitrary functions 
\begin{align*}
\phi_1=&-(k_1u+k_2v)-\omega_1, &\phi_3=-(k_1u+k_2v)-\omega_2,
\end{align*}
%we find Eq. \eqref{uvgen} reduces to the form
%\begin{subequations}
%\label{gMEE}
%\begin{align}
%\ddot{u}-(\alpha+1)h\dot{u}-\dot{h}u+[\alpha h^2+\omega_1(\alpha-1)h-\omega_1^2]u=0,  \\
%\ddot{v}-(\beta+1)h\dot{v}-\dot{h}v+[\beta h^2+\omega_2(\beta-1)h-\omega_2^2]v=0. 
%\end{align}
%\end{subequations}
we find \eqref{uvgen} reduces to the two-coupled version of the modified Emden equation \eqref{cMEE}, studied in the literature \cite{cMEE,nMEE}.  Substituting these forms of $\psi_1,\psi_2$ and $h$ in (\ref{cplsolnn}) we get the following particular solution of \eqref{cMEE} 
\begin{subequations}
\label{uvsoln1}
\begin{align}
u=\frac{C_1\omega_1\omega_2e^{\omega_1t}}{C_2\omega_1\omega_2+k_2\omega_1e^{\omega_2t}+C_1k_1\omega_2e^{\omega_1t}},  \\
v=\frac{\omega_1\omega_2e^{\omega_2t}}{C_2\omega_1\omega_2+k_2\omega_1e^{\omega_2t}+C_1k_1\omega_2e^{\omega_1t}},
\end{align}
\end{subequations}
where $C_1$ and $C_2$ are arbitrary constants.  One can check that this particular solution can be obtained from the general solution (13) given in Ref. \cite{nMEE} after fixing two of the integration constants.

We are also exploring further how the forms (\ref{redgen}) can be generalized so that more general system belonging to the class (\ref{cpl}) can be brought into the above formalism.  The results will be presented in future.

\section{Method of constructing general solution}
\label{sec4.5}
In the previous section we have obtained the particular solutions for a class of coupled second order ODEs by factorizing them.  In this section we obtain the general solution of a subset of  these coupled nonlinear ODEs (\ref{cplfact1}) by factorizing them for suitable parametric choices.  For this purpose we assume $\phi_i=\phi+\chi_i,\;i=1,2,3,4,$ where $\chi_i, \;i=1,2,3,4$, are arbitrary constants and $\phi$ is some arbitrary function of $u$ and $v$.  Then the equation (\ref{solgen}) becomes 
\begin{subequations}
\label{solgen1}
\begin{align}
[D-\phi(u,v)-\chi_1]P_1(u,v)&=0,\\
[D-\phi(u,v)-\chi_3]P_2(u,v)&=0,\\
[D-\phi(u,v)-\chi_2]\psi_1(u,v)&=P_1(u,v),\label{solgen1c}\\
[D-\phi(u,v)-\chi_4]\psi_2(u,v)&=P_2(u,v).\label{solgen1d}
\end{align}
\end{subequations}
This implies that the following relation holds good :
\begin{eqnarray}
\label{mod06}
 D\bigg(\frac{\psi_j}{P_j}\bigg)=\bar{\chi}_j\frac{\psi_j}{P_j}+1,\;j=1,2
\end{eqnarray}
where $\bar{\chi}_1=(\chi_2-\chi_1)$ and $\bar{\chi}_2=(\chi_4-\chi_3)$. Upon integrating equation (\ref{mod06}) we find a solution of the form
\begin{eqnarray}
\label{mod07}
 \frac{\psi_j}{P_j}=e^{\bar{\chi}_j t}\bigg(I_j-\frac{1}{\bar{\chi}_j}e^{-\bar{\chi}_j t}\bigg),\quad j=1,2
\end{eqnarray}
where $I_1$ and $I_2$ are two integration constants. Using \eqref{mod07}, one can rewrite Eqs. \eqref{solgen1c} and \eqref{solgen1d}  as
\begin{eqnarray}
\label{mod08}
 \frac{P_j}{\psi_j}=\frac{D[\psi_j(u,v)]}{\psi_j}-\phi(u,v)-\hat{\chi}_j=\frac{e^{-\bar{\chi}_j t}}{\bigg(I_j-\frac{1}{\bar{\chi}_j}e^{-\bar{\chi}_j t}\bigg)},\quad j=1,2
\end{eqnarray}
where $\hat{\chi}_1=\chi_2$ and $\hat{\chi}_2=\chi_4$. From (\ref{mod08}) we get
\begin{eqnarray}
\label{mod09}
 \frac{D[\psi_1]}{\psi_1}-\frac{D[\psi_2]}{\psi_2}-(\hat{\chi}_1-\hat{\chi}_2)=\frac{d}{dt}\bigg(\log\bigg[\frac{I_2-\frac{1}{\bar{\chi}_2}e^{-\bar{\chi}_2 t}}{I_1-\frac{1}{\bar{\chi}_1}e^{-\bar{\chi}_1 t}}\bigg]\bigg).
\end{eqnarray}
Integrating equation (\ref{mod09}) we obtain
\begin{eqnarray}
\label{mod10}
 \psi_2=\bigg[\frac{I_1-\frac{1}{\bar{\chi}_1}e^{-\bar{\chi}_1 t}}{I_2-\frac{1}{\bar{\chi}_2}e^{-\bar{\chi}_2 t}}\bigg]e^{(\hat{\chi}_2-\hat{\chi}_1)t+I_3}\psi_1=a(t)\psi_1,
\end{eqnarray}
where
\begin{eqnarray}
&&a(t)=\bigg[\frac{I_1-\frac{1}{\bar{\chi}_1}e^{-\bar{\chi}_1 t}}{I_2-\frac{1}{\bar{\chi}_2}e^{-\bar{\chi}_2 t}}\bigg]e^{(\hat{\chi}_2-\hat{\chi}_1)t+I_3}.\nonumber
\end{eqnarray}
Let us assume that the function $\phi(u,v)=h(\psi_1,\psi_2)=h(u,v)$, where $h(\psi_1,\psi_2)$ is given in (\ref{hdef3}). Now the equation for $\psi_1$ (see Eq. (\ref{mod08})) becomes
\begin{align}
\dot{\psi_1}=\bigg(\hat{\chi}_1-\frac{d}{dt}\bigg(\log\bigg[I_1-\frac{1}{\bar{\chi}_1}e^{-\bar{\chi}_1 t}\bigg]\bigg)\bigg)\psi_1 +s(t)\psi_1^{m+1},\label{bern1}
\end{align}
where \[s(t)=\sum_{i=1}^N k_ia(t)^{q_i},\qquad m=p_i+q_i,\] 
$k_0,k_1,\ldots,k_N$ are arbitrary parameters.

Equation \eqref{bern1} has the explicit solution 
\begin{subequations}
\label{sol112}
\begin{align}
\psi_1(u,v)=\frac{\bigg[I_1-\frac{1}{\bar{\chi}_1}e^{-\bar{\chi}_1 t}\bigg]e^{-m\hat{\chi}_1t}}{\left[I_4-m \int \bigg[I_1-\frac{1}{\bar{\chi}_1}e^{-\bar{\chi}_1 t}\bigg] e^{m\hat{\chi}_1t}s(t) dt \right]^{\frac{1}{m}}},\label{sol11}
\end{align}
\begin{align}
\psi_2(u,v)=\frac{a(t)\bigg[I_1-\frac{1}{\bar{\chi}_1}e^{-\bar{\chi}_1 t}\bigg]e^{-m\hat{\chi}_1t}}{\left[I_4-m \int \bigg[I_1-\frac{1}{\bar{\chi}_1}e^{-\bar{\chi}_1 t}\bigg] e^{m\hat{\chi}_1t}s(t) dt \right]^{\frac{1}{m}}}.\label{sol12}
\end{align}
\end{subequations}
From the relations (\ref{psidef}) we find
\begin{eqnarray}
u=\frac{b_2}{\delta}(\psi_1-c_1)-\frac{b_1}{\delta}(\psi_2-c_2),\\
v=\frac{a_1}{\delta}(\psi_2-c_2)-\frac{a_2}{\delta}(\psi_1-c_1),
\end{eqnarray}
where $\psi_1$ and $\psi_2$ are given by Eq. (\ref{sol112}), $\delta=a_1b_2-a_2b_1$.
Note that the above solution contains four arbitrary constants $I_1,\,I_2,\,I_3,$ and $I_4$, (with $I_2,\,I_3$ appearing in the function $a(t)$, see Eq. (\ref{mod10})) so that (\ref{sol11})-(\ref{sol12}) constitute the general solution.
\subsection{Example}
We illustrate the above procedure by considering the following equation belonging to the class of equations given by (\ref{example1}),
\begin{subequations}
\bea
&&\hspace{-0.5cm}\ddot{u}+k_2u\dot{v}+(3k_1u+2k_2v-\chi_1-\chi_2)\dot{u}+k_2uv(k_2v-\chi_1-\chi_2)
-k_1u^2(\chi_1+\chi_2-2k_2v)\nonumber\\
&&\hspace{3cm}+k_1^2u^3+\chi_1\chi_2u=0,\\
&&\hspace{-0.5cm}\ddot{v}+k_1v\dot{u}+(3k_2v+2k_1u-\chi_3-\chi_4)\dot{v}+k_1uv(k_1u-\chi_3-\chi_4)
-k_2v^2(\chi_3+\chi_4-2k_1u)\nonumber\\
&&\hspace{3cm}+k_2^2v^3+\chi_3\chi_4v=0,
\eea
\end{subequations}
which can be factorized as
%with $\psi_1(u,v)=u$, $\psi_2(u,v)=v$ and $\phi(u,v)=-(k_1u+k_2v)$ in Eq. ()
\begin{subequations}
\label{eq-eq}
\bea
&&\left[D+(k_1u+k_2v)-\chi_1\right]P_1(u,v)=0,\\
&&\left[D+(k_1u+k_2v)-\chi_3\right]P_2(u,v)=0,\\
&&\left[D+(k_1u+k_2v)-\chi_2\right]u=P_1(u,v),\label{eg-eq-c}\\
&&\left[D+(k_1u+k_2v)-\chi_4\right]v=P_2(u,v).\label{eg-eq-d}
\eea
\end{subequations}
We obtain the following relations from the above equations,
\begin{subequations}
\bea
&&D\left(\frac{u}{P_1}\right)=\bar{\chi_1}\frac{u}{P_1}+1,\\
&&D\left(\frac{v}{P_2}\right)=\bar{\chi_2}\frac{v}{P_2}+1,
\eea
\end{subequations}
where $\bar{\chi_1}=\chi_2-\chi_1$ and $\bar{\chi_2}=\chi_4-\chi_3$.
Integrating we get
\bea
 \frac{u}{P_1}=e^{\bar{\chi}_1 t}\bigg(I_1-\frac{1}{\bar{\chi}_1}e^{-\bar{\chi}_1 t}\bigg),\qquad \frac{v}{P_2}=e^{\bar{\chi}_2 t}\bigg(I_2-\frac{1}{\bar{\chi}_2}e^{-\bar{\chi}_2 t}\bigg).
\eea
Using these relations in (\ref{eg-eq-c}) and (\ref{eg-eq-d}) we get
\begin{subequations}
\bea
\frac{\dot{u}}{u}+(k_1u+k_2v)-\bar{\chi}_1=\frac{e^{-\bar{\chi}_1t}}
{\left(I_1-\frac{1}{\bar{\chi}_j}e^{-\bar{\chi}_1t}\right)},\label{eg-eq-u}\\
\frac{\dot{v}}{v}+(k_1u+k_2v)-\bar{\chi}_2=\frac{e^{-\bar{\chi}_2t}}
{\left(I_2-\frac{1}{\bar{\chi}_2}e^{-\bar{\chi}_2t}\right)}.
\eea
\end{subequations}
Integrating the above system of equations we get
\bea
v=\bigg[\frac{I_1-\frac{1}{\bar{\chi}_1}e^{-\bar{\chi}_1 t}}{I_2-\frac{1}{\bar{\chi}_2}e^{-\bar{\chi}_2 t}}\bigg]e^{(\hat{\chi}_2-\hat{\chi}_1)t+I_3}u=a(t)u\label{eg-eq-v}
\eea
Substituting for $v$ in \eqref{eg-eq-u} we get
\bea
\dot{u}=\hat{\chi}_1u-\frac{d}{dt}\left(\log\left[I_1-\frac{1}{\bar{\chi}_1}e^{-\bar{\chi}_1t}\right]\right)u+(k_1+k_2a(t))u^2.
\eea
We wish to note that the above equation falls under the Riccati equation which upon integration leads to the following general solution,
\bea
u= \frac{\left(I_1-\frac{1}{\bar{\chi}_1}e^{-\bar{\chi}_1t}\right)e^{\bar{\chi}_1t}}{\bigg[I_4+\int \left(I_1-\frac{1}{\bar{\chi}_1}e^{-\bar{\chi}_1t}\right)e^{\bar{\chi}_1t}(k_1+k_2a(t))dt\bigg]}.
\eea
Substituting for $u$ in \eqref{eg-eq-v} we get
\bea
v=\frac{a(t)\left(I_1-\frac{1}{\bar{\chi}_1}e^{-\bar{\chi}_1t}\right)e^{\bar{\chi}_1t}}
{\bigg[I_4+\int \left(I_1-\frac{1}{\bar{\chi}_1}e^{-\bar{\chi}_1t}\right) e^{\bar{\chi}_1t}(k_1+k_2a(t))dt\bigg] }.
\eea
We note here that the above solution contains four integration constants $I_1,\,I_2,\,I_3$ and $I_4$ and this solution can also be obtained by following the procedure discussed in Ref. \cite{nonlocal}.  We find that the general solution of a restricted class of equations, a subset of the class of equations  (\ref{example1}) ,  which can be factorized in the form (\ref{solgen1}) can be obtained using the above procedure.  For the case of scalar second order ODEs discussed in Ref. \cite{cornejo2}, obviously we can apply the above procedure straightforwardly.  The details are given in Appendix C.
It should be possible to generalize the above procedure to more general cases than (\ref{solgen1}), though we do not attempt this in this paper.

\section{Conclusion}
\label{sec6}
In this paper we have identified a system of coupled Li\'enard type equations which can be factorized in terms of first order differential operators.  We have shown that a particular solution of equations belonging to this class of coupled Li\'enard type equations can be obtained by solving a Bernoulli type equation. This generic class of ODEs contains the a coupled version of the modified Emden equation  which has been recently studied in the literature. We have also shown that the general solution of a restricted class of equations can also be obtained using this procedure of factorization.  Several generalizations of our study can be proceeded with, by relaxing the various restrictions mentioned in the present work.  These are being pursued currently.

In addition to this, one can extend this procedure to higher order scalar/coupled ODEs and obtain their corresponding particular/general solutions for suitable choice of the parameters.  One can also straightforwardly extend this procedure of factorization to a system of $N$ coupled second order ODEs.  The factorization of $N$ coupled second order ODEs will result in a system of $2N$ coupled first order ODEs similar to (\ref{solgen1}).  From these $2N$ first order ODEs one can obtain a relation similar to Eq. (\ref{mod06}) with $j=1,2,\ldots,N$.  Integrating this system of equations one can obtain the general solution of the underlying $N$ coupled second order ODEs.  

\section{Acknowledgments}

The work of TH is part of Summer Research Fellowship Programme by the IASc-INSA-NASI. TH is supported by a KVPY Fellowship sponsored by a Department of Science and Technology (DST), Government of India. The work of VKC, RGP and ML is supported by a DST--IRHPA research project. The work of ML is also supported by DST--Ramanna Fellowship program and a DAE Raja Ramanna Fellowship.

\appendix 
\section{Proof of $\psi_{1,2uu}=\psi_{1,2uv}=\psi_{1,2vv}=0$}
\label{app1}

Requiring that the coefficients of the higher powers of the derivatives contained in the bracketed terms in \eqref{cplfact} be zero we get
%\begin{subequations}
\begin{align}
(\psi_{1uu}\psi_{2v}-\psi_{2uu}\psi_{1v})\dot{u}^2+(\psi_{1uv}\psi_{2v}-\psi_{2uv}\psi_{1v})\dot{u}\dot{v}+(\psi_{1vv}\psi_{2v}-\psi_{2vv}\psi_{1v})\dot{v}^2=0 \\
(\psi_{1uu}\psi_{2u}-\psi_{2uu}\psi_{1u})\dot{u}^2+(\psi_{1uv}\psi_{2u}-\psi_{2uv}\psi_{1u})\dot{u}\dot{v}+(\psi_{1vv}\psi_{2u}-\psi_{2vv}\psi_{1u})\dot{v}^2=0 
\end{align}
%\end{subequations}
Since $u$ and $v$ are independent the coefficients of $\dot{u}^2,\dot{u}\dot{v},\dot{v}^2$ in each equation must individually be zero.
\begin{align}
\Rightarrow \qquad \qquad \qquad&\psi_{1uu}\psi_{2v}=\psi_{2uu}\psi_{1v}, &\psi_{1uu}\psi_{2u}=\psi_{2uu}\psi_{1u}, \notag \\
 &\psi_{1uv}\psi_{2v}=\psi_{2uv}\psi_{1v}, &\psi_{1uv}\psi_{2u}=\psi_{2uv}\psi_{1u},\notag \\
 &\psi_{1vv}\psi_{2v}=\psi_{2vv}\psi_{1v}, &\psi_{1vv}\psi_{2u}=\psi_{2vv}\psi_{1u},
\label{psi1}
\end{align}
If $\psi_{2uu},\psi_{2uv},\psi_{2vv}\neq 0$ then \eqref{psi1} implies
\begin{align}
\frac{\psi_{1uu}}{\psi_{2uu}}=\frac{\psi_{1uv}}{\psi_{2uv}}=\frac{\psi_{1vv}}{\psi_{2vv}}=\frac{\psi_{1u}}{\psi_{2u}}=\frac{\psi_{1v}}{\psi_{2v}}\label{psi2}
\end{align}
The last equality in \eqref{psi2} implies $\delta=\psi_{1u}\psi_{2v}-\psi_{2u}\psi_{1v}=0$ which is not permissible in \eqref{cplfact} as in that case the leading order terms $\ddot{u},\ddot{v}$ vanish and the resulting equation is first order. Thus $\psi_{2uu},\psi_{2uv},\psi_{2vv}=0$.

If $\psi_{2uu},\psi_{2uv},\psi_{2vv}= 0 \mbox{ but one of } \psi_{1uu},\psi_{1uv},\psi_{1vv}\neq 0$ then from Eq \eqref{psi2} $\psi_{2u}=\psi_{2v}=0$ which again implies $\delta=0$ which is inadmissible. Thus  $\psi_{1,2uu}=\psi_{1,2uv}=\psi_{1,2vv}=0$.

\section{Separable reduced equations}
\label{app2}
We consider the case where $h_1$ and $h_2$ in (\ref{redgen1}-\ref{redgen2}) are of the form
\begin{align}
h_1=k_1\psi_1^p,\: h_2=k_2\psi_2^q.\label{hsep}
\end{align}
The reduced equations are then a pair of Bernoulli equations in $\psi_1$ and $\psi_2$
\begin{align}
\dot{\psi_1}=\omega_1\psi_1+k_1\psi_1^{p+1},\notag\\
\dot{\psi_2}=\omega_2\psi_2+k_2\psi_2^{q+1}.\label{sepbern}
\end{align}

For the class of systems for which \eqref{expred} is equivalent to (\ref{redgen1}-\ref{redgen2}) with $h_1$ and $h_2$ given by \eqref{hsep} explicit solutions can be found relatively simply by solving \eqref{sepbern}. Using the condition $\phi_2=\omega_1+k_1\psi_1^p$ and $\phi_4=\omega_2+k_2\psi_2^q$ and \eqref{phi13gen2} in \eqref{cplfact}, we identify this class of systems to be
\begin{subequations}
\label{sepclass}
\begin{align}
&&\hspace{-1cm}\ddot{u}+\frac{1}{\delta}\bigg[(a_2b_1\bar{g_2}+\omega_2+k_2(q+1)\psi_2^q)-a_1b_2(\bar{g_1}+\omega_1+k_1(p+1)\psi_1^p)\bigg]\dot{u} \notag\\
        &&+\frac{1}{\delta}\bigg[b_1b_2(\bar{g_2}-\bar{g_1}+\omega_2-\omega_1+k_2(q+1)\psi_2^q-k_1(p+1)\psi_1^p)\bigg]\dot{v} \notag\\
        &&+\frac{1}{\delta}\bigg[b_2\bar{g_1}(\omega_1+k_1\psi_1^p)\psi_1-b_1\bar{g_2}(\omega_2+k_2\psi_2^q)\psi_2\bigg]=0, \\
&&\hspace{-1cm}\ddot{v}-\frac{1}{\delta}\bigg[a_1b_2(\bar{g_2}+\omega_2+k_2(q+1)\psi_2^q)-a_2b_1(\bar{g_1}+\omega_1+k_1(p+1)\psi_2^p)\bigg]\dot{v} \notag\\
        &&-\frac{1}{\delta}\bigg[a_1a_2(\bar{g_2}-\bar{g_1}+\omega_2-\omega_1+k_2(q+1)\psi_2^q-k_1(p+1)\psi_1^p)\bigg]\dot{u} \notag\\
        &&-\frac{1}{\delta}\bigg[a_2\bar{g_1}(\omega_1+k_1\psi_1^p)\psi_1-a_1\bar{g_2}(\omega_2+k_2\psi_2^q)\psi_2\bigg]=0,
\end{align}
\end{subequations}
where $\psi_1$ and $\psi_2$ are given by \eqref{psidef} and $\bar{g_1}$ and $\bar{g_2}$ are given by \eqref{Gdef}. 

The solution to \eqref{sepbern} is found using (\ref{sol1}-\ref{sol2}) to be 
\begin{align}
\psi_1=\left[k_1(C_1e^{-p\omega_1t}-\frac{p}{\omega_1})\right]^{-\frac{1}{p}},\notag\\
\psi_2=\left[k_2(C_2e^{-q\omega_2t}-\frac{q}{\omega_2})\right]^{-\frac{1}{q}}.\notag
\end{align}
Using \eqref{inv} we get an explicit solution for $u$ and $v$ as
\begin{align}
u=\phantom{-}\frac{b_2}{\delta}\left(\left[k_1(C_1e^{-p\omega_1t}-\frac{p}{\omega_1})\right]^{-\frac{1}{p}}-c_1\right)-\frac{b_1}{\delta}\left(\left[k_2(C_2e^{-q\omega_2t}-\frac{q}{\omega_2})\right]^{-\frac{1}{q}}-c_2\right),\notag\\
v=-\frac{a_2}{\delta}\left(\left[k_1(C_1e^{-p\omega_1t}-\frac{p}{\omega_1})\right]^{-\frac{1}{p}}-c_1\right)+\frac{a_1}{\delta}\left(\left[k_2(C_2e^{-q\omega_2t}-\frac{q}{\omega_2})\right]^{-\frac{1}{q}}-c_2\right),\notag
\end{align}
where $C_1$ and $C_2$ are integration constants.
\section{General solution of scalar ODEs}
Using the factorization procedure Cornejo-P\'erez et al. in Ref. \cite{cornejo2} have obtained particular solutions of a class of scalar second order ODEs.  Now, we show that the procedure of obtaining general solution discussed in Sec. \ref{sec4.5} is also applicable to scalar second order ODEs.  We demonstrate this by considering a specific equation discussed in Ref. \cite{cornejo2} and obtain its general solution through factorization for suitable parametric choice.
Let us consider the following equation
\begin{eqnarray}
\ddot{x}+(a-\alpha x^p)\dot{x}+\beta x(1-x^p)(x^p-b)=0,\label{scalar1}
\end{eqnarray}
which is obtained from the Burgers-Huxley equation \cite{cornejo2} using a traveling wave reduction.  We find that for the parametric choice $\beta=\frac{\alpha^2}{(p+2)^2},\,\,b=-\frac{a}{\alpha}(p+2)-1,\,\,\alpha=-\frac{1}{2}(p+2)(a\pm(\chi_2-\chi_1)$, Eq. (\ref{scalar1}) can be factorized in the form
\begin{subequations}
\label{scalar-factor}
\begin{eqnarray}
(D-\phi-\chi_1)P=0,\\
(D-\phi-\chi_2)x=P,\label{scalar-factor2}
\end{eqnarray}
\end{subequations}
where $D=\frac{d}{dt}$, $\phi=\frac{\alpha}{(p+2)}x^p-\frac{1}{2}(\chi_1+\chi_2+a)$, $\chi_1$ and $\chi_2$ are arbitrary parameters.  From Eq. (\ref{scalar-factor}) and following the procedure discussed in Sec. \ref{sec4.5} we get
\begin{eqnarray}
D\left(\frac{x}{P}\right)=(\chi_2-\chi_1)\frac{x}{P}+1.
\end{eqnarray}
Integrating we obtain
\begin{eqnarray}
\frac{x}{P}=e^{(\chi_2-\chi_1)t}\left(I_1-\frac{1}{(\chi_2-\chi_1)}e^{-(\chi_2-\chi_1)t}\right).
\end{eqnarray}
Rewriting Eq. (\ref{scalar-factor}) and substituting for $\frac{P}{x}$ from the above equation we get
\begin{eqnarray}
\frac{D[x]}{x}-\phi-\chi_2=\frac{e^{(\chi_1-\chi_2)t}}
{\left(I_1-\frac{1}{(\chi_2-\chi_1)}e^{(\chi_1-\chi_2)t}\right)},
\end{eqnarray}
where $\phi=\frac{\alpha}{(p+2)}x^p-\frac{1}{2}(\chi_1+\chi_2+a)$.  Integrating the above Bernoulli type equation we obtain
\begin{eqnarray}
x(t)=\frac{\left(I_1-\frac{e^{(\chi_1-\chi_2)t}}{\chi_2-\chi_1}\right)e^{-\frac{1}{2}(\chi_2-\chi_1+a)t}}{\left(I_2-p\int \left(I_1-\frac{e^{(\chi_1-\chi_2)t}}{\chi_2-\chi_1}\right)^pe^{-\frac{p}{2}(\chi_2-\chi_1+a)t} dt\right)^p}
\end{eqnarray}
Similarly one can obtain the general solution for all the other equations discussed in Ref. \cite{cornejo2} for suitable choice of parameters using the procedure discussed here.

%\bibliography{test.bib}

\begin{thebibliography}{22}
\expandafter\ifx\csname natexlab\endcsname\relax\def\natexlab#1{#1}\fi
\expandafter\ifx\csname bibnamefont\endcsname\relax
  \def\bibnamefont#1{#1}\fi
\expandafter\ifx\csname bibfnamefont\endcsname\relax
  \def\bibfnamefont#1{#1}\fi
\expandafter\ifx\csname citenamefont\endcsname\relax
  \def\citenamefont#1{#1}\fi
\expandafter\ifx\csname url\endcsname\relax
  \def\url#1{\texttt{#1}}\fi
\expandafter\ifx\csname urlprefix\endcsname\relax\def\urlprefix{URL }\fi
\providecommand{\bibinfo}[2]{#2}
\providecommand{\eprint}[2][]{\url{#2}}

\bibitem[{\citenamefont{Painlev\'e}(1902)}]{painleve}
\bibinfo{author}{\bibfnamefont{P.}~\bibnamefont{Painlev\'e}},
  \bibinfo{journal}{Acta Math.} \textbf{\bibinfo{volume}{25}},
  \bibinfo{pages}{1} (\bibinfo{year}{1902}).

\bibitem[{\citenamefont{Chandrasekar et~al.}(2007)\citenamefont{Chandrasekar,
  Senthilvelan, and Lakshmanan}}]{gensoln}
\bibinfo{author}{\bibfnamefont{V.~K.} \bibnamefont{Chandrasekar}},
  \bibinfo{author}{\bibfnamefont{M.}~\bibnamefont{Senthilvelan}},
  \bibnamefont{and}
  \bibinfo{author}{\bibfnamefont{M.}~\bibnamefont{Lakshmanan}},
  \bibinfo{journal}{J. Phys. A: Math. Theor.} \textbf{\bibinfo{volume}{40}},
  \bibinfo{pages}{4717} (\bibinfo{year}{2007}).

\bibitem[{\citenamefont{Moreira}(1984)}]{gas11}
\bibinfo{author}{\bibfnamefont{I.~C.} \bibnamefont{Moreira}},
  \bibinfo{journal}{Hadronic. J.} \textbf{\bibinfo{volume}{7}},
  \bibinfo{pages}{475} (\bibinfo{year}{1984}).

\bibitem[{\citenamefont{Leach}(1985)}]{gas12}
\bibinfo{author}{\bibfnamefont{P.~G.~L.} \bibnamefont{Leach}},
  \bibinfo{journal}{J. Math. Phys.} \textbf{\bibinfo{volume}{26}},
  \bibinfo{pages}{2510} (\bibinfo{year}{1985}).

\bibitem[{\citenamefont{Chandrasekhar}(1957)}]{gas21}
\bibinfo{author}{\bibfnamefont{S.}~\bibnamefont{Chandrasekhar}},
  \emph{\bibinfo{title}{An Introduction to the Study of Stellar Structure}}
  (\bibinfo{publisher}{Dover}, \bibinfo{address}{New York},
  \bibinfo{year}{1957}).

\bibitem[{\citenamefont{Dixon and Tuszynski}(1990)}]{gas22}
\bibinfo{author}{\bibfnamefont{J.~M.} \bibnamefont{Dixon}} \bibnamefont{and}
  \bibinfo{author}{\bibfnamefont{J.~A.} \bibnamefont{Tuszynski}},
  \bibinfo{journal}{Phys. Rev. A} \textbf{\bibinfo{volume}{41}},
  \bibinfo{pages}{4166} (\bibinfo{year}{1990}).

\bibitem[{\citenamefont{Erwin et~al.}(1984)\citenamefont{Erwin, Ames, and
  Adams}}]{fusion}
\bibinfo{author}{\bibfnamefont{V.~J.} \bibnamefont{Erwin}},
  \bibinfo{author}{\bibfnamefont{W.~F.} \bibnamefont{Ames}}, \bibnamefont{and}
  \bibinfo{author}{\bibfnamefont{E.}~\bibnamefont{Adams}},
  \emph{\bibinfo{title}{Wave Phenomena: Modern Theory and Applications}}
  (\bibinfo{address}{Amsterdam: North-Holland}, \bibinfo{year}{1984}).

\bibitem[{\citenamefont{McVittie}(1933)}]{stellar1}
\bibinfo{author}{\bibfnamefont{G.~C.} \bibnamefont{McVittie}},
  \bibinfo{journal}{Mon. Not. R. Astron. Soc.} \textbf{\bibinfo{volume}{93}},
  \bibinfo{pages}{325} (\bibinfo{year}{1933}).

\bibitem[{\citenamefont{McVittie}(1967)}]{stellar2}
\bibinfo{author}{\bibfnamefont{G.~C.} \bibnamefont{McVittie}},
  \bibinfo{journal}{Ann. Inst. H Poincar\'e} \textbf{\bibinfo{volume}{6}},
  \bibinfo{pages}{1} (\bibinfo{year}{1967}).

\bibitem[{\citenamefont{McVittie}(1984)}]{stellar3}
\bibinfo{author}{\bibfnamefont{G.~C.} \bibnamefont{McVittie}},
  \bibinfo{journal}{Ann. Inst. H Poincar\'e} \textbf{\bibinfo{volume}{3}},
  \bibinfo{pages}{231} (\bibinfo{year}{1984}).

\bibitem[{\citenamefont{Chisholm and Common}(1987)}]{boson1}
\bibinfo{author}{\bibfnamefont{J.~S.~R.} \bibnamefont{Chisholm}}
  \bibnamefont{and} \bibinfo{author}{\bibfnamefont{A.~K.}
  \bibnamefont{Common}}, \bibinfo{journal}{J. Phys. A: Math. Gen.}
  \textbf{\bibinfo{volume}{20}}, \bibinfo{pages}{5459} (\bibinfo{year}{1987}).

\bibitem[{\citenamefont{Yang and Mills}(1954)}]{boson2}
\bibinfo{author}{\bibfnamefont{C.~N.} \bibnamefont{Yang}} \bibnamefont{and}
  \bibinfo{author}{\bibfnamefont{R.~L.} \bibnamefont{Mills}},
  \bibinfo{journal}{Phys. Rev.} \textbf{\bibinfo{volume}{96}},
  \bibinfo{pages}{191} (\bibinfo{year}{1954}).

\bibitem[{\citenamefont{Golubev}(1950)}]{math1}
\bibinfo{author}{\bibfnamefont{V.~V.} \bibnamefont{Golubev}},
  \emph{\bibinfo{title}{Lectures on Analytical Theory of Differential
  Equations}} (\bibinfo{publisher}{Gostekhizdat}, \bibinfo{address}{Moscow},
  \bibinfo{year}{1950}).

\bibitem[{\citenamefont{van der Pol}(1927)}]{idunno1}
\bibinfo{author}{\bibfnamefont{B.} \bibnamefont{van der Pol}},
  \bibinfo{journal}{Philos. Mag.} \textbf{\bibinfo{volume}{3}},
  \bibinfo{pages}{65} (\bibinfo{year}{1927}).

\bibitem[{\citenamefont{Moreira}(1992)}]{idunno2}
\bibinfo{author}{\bibfnamefont{H.~N.} \bibnamefont{Moreira}},
  \bibinfo{journal}{Ecological Modelling} \textbf{\bibinfo{volume}{60}},
  \bibinfo{pages}{139} (\bibinfo{year}{1992}).

\bibitem[{\citenamefont{Cornejo~P\'erez and Rosu}(2005)}]{cornejo2}
\bibinfo{author}{\bibfnamefont{O.}~\bibnamefont{Cornejo~P\'erez}}
  \bibnamefont{and} \bibinfo{author}{\bibfnamefont{H.~C.} \bibnamefont{Rosu}},
  \bibinfo{journal}{Prog. Theor. Phys.} \textbf{\bibinfo{volume}{114}},
  \bibinfo{pages}{533} (\bibinfo{year}{2005}).

\bibitem[{\citenamefont{Cornejo~P\'erez and Rosu}(2006)}]{cornejo3}
\bibinfo{author}{\bibfnamefont{O.}~\bibnamefont{Cornejo~P\'erez}}, \bibinfo{author}{\bibfnamefont{J.} \bibnamefont{Negro}},
\bibinfo{author}{\bibfnamefont{L.M} \bibnamefont{Nieto}}
  \bibnamefont{and} \bibinfo{author}{\bibfnamefont{H.~C.} \bibnamefont{Rosu}},
  \bibinfo{journal}{Found. Phys.} \textbf{\bibinfo{volume}{36}},
  \bibinfo{pages}{1587} (\bibinfo{year}{2006}).


\bibitem[{\citenamefont{Berkovich}(2007)}]{berk}
\bibinfo{author}{\bibfnamefont{L.~M.} \bibnamefont{Berkovich}},
  \bibinfo{journal}{Applicable Analysis and Discrete Mathematics}
  \textbf{\bibinfo{volume}{1}}, \bibinfo{pages}{122} (\bibinfo{year}{2007}).

\bibitem[{\citenamefont{Berkovich}(2007)}]{li}
\bibinfo{author}{\bibfnamefont{D.~S.} \bibnamefont{Wang}}
  \bibnamefont{and} \bibinfo{author}{\bibfnamefont{H.} \bibnamefont{Li}},
  \bibinfo{journal}{J. Math. Anal. Appl}
  \textbf{\bibinfo{volume}{343}}, \bibinfo{pages}{273} (\bibinfo{year}{2008}).


\bibitem[{\citenamefont{Reyes and Rosu}(2008)}]{scfact}
\bibinfo{author}{\bibfnamefont{M.~A.} \bibnamefont{Reyes}} \bibnamefont{and}
  \bibinfo{author}{\bibfnamefont{H.~C.} \bibnamefont{Rosu}},
  \bibinfo{journal}{J. Phys. A: Math. Theor.} \textbf{\bibinfo{volume}{41}},
  \bibinfo{pages}{285206} (\bibinfo{year}{2008}).

\bibitem[{\citenamefont{Gladwin~Pradeep
  et~al.}(2009)\citenamefont{Gladwin~Pradeep, Chandrasekar, Senthilvelan, and
  Lakshmanan}}]{nMEE}
\bibinfo{author}{\bibfnamefont{R.}~\bibnamefont{Gladwin~Pradeep}},
  \bibinfo{author}{\bibfnamefont{V.~K.} \bibnamefont{Chandrasekar}},
  \bibinfo{author}{\bibfnamefont{M.}~\bibnamefont{Senthilvelan}},
  \bibnamefont{and}
  \bibinfo{author}{\bibfnamefont{M.}~\bibnamefont{Lakshmanan}},
  \bibinfo{journal}{J. Phys. A : Math. Theor.} \textbf{\bibinfo{volume}{42}},
  \bibinfo{pages}{135206} (\bibinfo{year}{2009}).

\bibitem[{\citenamefont{Chandrasekar et~al.}(2009)\citenamefont{Chandrasekar,
  Senthilvelan, and Lakshmanan}}]{cMEE}
\bibinfo{author}{\bibfnamefont{V.~K.} \bibnamefont{Chandrasekar}},
  \bibinfo{author}{\bibfnamefont{M.}~\bibnamefont{Senthilvelan}},
  \bibnamefont{and}
  \bibinfo{author}{\bibfnamefont{M.}~\bibnamefont{Lakshmanan}},
  \bibinfo{journal}{Proc. R. Soc. A} \textbf{\bibinfo{volume}{465}},
  \bibinfo{pages}{609} (\bibinfo{year}{2009}).

\bibitem[{\citenamefont{Chandrasekar et~al.}(2005)\citenamefont{Chandrasekar,
  Senthilvelan, and Lakshmanan}}]{isochr}
\bibinfo{author}{\bibfnamefont{V.~K.} \bibnamefont{Chandrasekar}},
  \bibinfo{author}{\bibfnamefont{M.}~\bibnamefont{Senthilvelan}},
  \bibnamefont{and}
  \bibinfo{author}{\bibfnamefont{M.}~\bibnamefont{Lakshmanan}},
  \bibinfo{journal}{Phys. Rev. E} \textbf{\bibinfo{volume}{72}},
  \bibinfo{pages}{066203} (\bibinfo{year}{2005}).

\bibitem[{\citenamefont{Gladwin~Pradeep
  et~al.}(2010)\citenamefont{Gladwin~Pradeep, Chandrasekar, Senthilvelan, and
  Lakshmanan}}]{nonlocal}
\bibinfo{author}{\bibfnamefont{R.}~\bibnamefont{Gladwin~Pradeep}},
  \bibinfo{author}{\bibfnamefont{V.~K.} \bibnamefont{Chandrasekar}},
  \bibinfo{author}{\bibfnamefont{M.}~\bibnamefont{Senthilvelan}},
  \bibnamefont{and}
  \bibinfo{author}{\bibfnamefont{M.}~\bibnamefont{Lakshmanan}},
  \bibinfo{journal}{J. Math. Phys.} \textbf{\bibinfo{volume}{51}},
  \bibinfo{pages}{103513} (\bibinfo{year}{2010}).

\end{thebibliography}
%\include{pra.bbl}
\end{document}